\documentclass[12pt]{article}
\renewcommand{\baselinestretch}{1.5}

\usepackage{subfigure,float}

\usepackage{booktabs}
\usepackage{amsmath,amssymb}
\usepackage{color}
\usepackage{hyperref}
\usepackage{latexsym}
\usepackage{graphicx,graphics}
\usepackage{tabu}
\usepackage{longtable}
\usepackage{natbib}
\DeclareGraphicsExtensions{.bmp,.jpg,.eps,.pdf,.png}
\usepackage{float,verbatim}

\usepackage{mathtools,amssymb,amsthm}
\usepackage{bm}
\usepackage[linesnumbered,ruled,vlined]{algorithm2e}
\newcommand{\bB}{\bm{B}}

\newcommand{\bI}{\bm{I}}

\newcommand{\bV}{\bm{V}}

\newcommand{\bx}{\bm{x}}

\newcommand{\bK}{\bm{K}}

\newcommand{\bY}{\bm{Y}}

\newcommand{\bk}{\bm{k}}

\newcommand{\br}{\bm{r}}
\newcommand{\bv}{\bm{v}}
\newcommand{\bu}{\bm{u}}

\newcommand{\btht}{\bm{\theta}}

\newcommand{\hbt}{\hat{\bm{\theta}}}
\newcommand{\hbT}{\hat{\bm{\Theta}}}

\newcommand{\hbc}{\hat{\bm{c}}}

\newcommand{\tvec}{\text{vec}}

\begin{document}
\title{The Evolution of Dynamic Gaussian Process Model with Applications to Malaria Vaccine Coverage Prediction}

\renewcommand{\baselinestretch}{1.0}
\author{Pritam Ranjan\footnote{Corresponding author: pritamr@iimidr.ac.in}, Indian Institute of Management Indore\\
{M. Harshvardhan, Indian Institute of Management Indore}\\}
%\author{*****\footnote{Corresponding author: ******}, ******\\
%	{********}\\}

\date{}

\maketitle

\begin{abstract}
Gaussian process (GP) based statistical surrogates are popular, inexpensive substitutes for emulating the outputs of expensive computer models that simulate  real-world phenomena or complex systems. Here, we discuss the evolution of dynamic GP model --- a computationally efficient statistical surrogate for a computer simulator with time series outputs. The main idea is to use a convolution of standard GP models, where the weights are guided by a singular value decomposition (SVD) of the response matrix over the time component. The dynamic GP model also adopts a localized modeling approach for building a statistical model for large datasets.

In this chapter, we use several popular test function based computer simulators to illustrate the evolution of dynamic GP models. We also use this model for predicting the coverage of Malaria vaccine worldwide. Malaria is still affecting more than eighty countries concentrated in the tropical belt. In 2019 alone, it was the cause of more than 435,000 deaths worldwide. The malice is easy to cure if diagnosed in time, but the common symptoms make it difficult. We focus on a recently discovered reliable vaccine called Mos-Quirix (RTS,S) which is currently going under human trials. With the help of publicly available data on dosages, efficacy, disease incidence and communicability of other vaccines obtained from the World Health Organisation, we predict vaccine coverage for 78 Malaria-prone countries. 
\end{abstract}

\section{Introduction}

Computer simulators are widely used to understand complex physical systems in many areas such as aerospace, renewable energy, climate modelling, and manufacturing. For example, \cite{greenberg1979numerical} developed a finite volume community ocean model (FVCOM) for simulating the flow of water in the Bay of Fundy; \cite{bower2006breaking} discussed the formation of galaxies using a simulator called GALFORM; \cite{bayarri2009using} used a simulator called TITAN2D for modelling the maximum volcanic eruption flow height; and \cite{zhang2017local} used a TDB simulator to model the population growth of European red mites. Realistic computer simulators can also be computationally expensive to run, and thus statistical surrogates used as an inexpensive substitute for a deeper understanding of the underlying phenomena. \cite{sacks1989} proposed using a realization  of the Gaussian process (GP) model as a surrogate for such simulator outputs.

The types of simulator outputs structures dealt with are as varied as the applications. One is faced with scalar, multivariate, functional, time series and spatial-temporal data, to name a few.  In this chapter, we discuss the evolution of  GP-based surrogate models for computer simulators with time series outputs, which we refer to as {\em dynamic computer simulators}.  Such simulators arise in various application, for example, rainfall-runoff model \citep{conti2009gaussian}, vehicle suspension system \citep{bayarri2007computer}, and TDB model \citep{zhang2017local}.

 The emulation of dynamic computer simulators has been considered by many \citep{kennedy2001bayesian, stein2005space, bayarri2007computer, higdon2008computer, conti2009gaussian, liu2009dynamic, farah2014bayesian, hung2015analysis}. In this chapter, we highlight the singular value decomposition (SVD)-based GP models, which was originally introduced by \citet{higdon2008computer}  for computer model calibration with high-dimensional outputs. However, \cite{zhang2017local} generalized it further for time-series responses and developed the empirical Bayesian inference for large-scale computer simulators.

Fitting GP models requires the inversion of $N \times N$ spatial correlation matrices, which gets prohibitive if $N$ (the sample size) becomes large. {In other words}, fitting GP models over the entire training set can often be computationally infeasible for large-scale dynamic computer experiments involving thousands of training points.
 A naive popular approach is to build localized models for prediction in the big data context. To search for the most relevant data for local neighborhood in a more intelligent way,  \citet{emery2009kriging} built a local neighborhood by sequentially including data that make the kriging variance decrease more.  \citet{gramacy2015local} improved the prediction accuracy by using a sequential greedy algorithm and an optimality criterion for finding a non-trivial local neighborhood set, and \cite{zhang2017local} further extended the idea for dynamic simulator outputs.

In this {chapter}, we illustrate the implementation of dynamic svd-based GP model for several test function based simulator outputs, and a real-life modeling problem where the objective is to predict the usage of a new Malaria vaccine. {Malaria is a mosquito-borne disease caused by a \textit{Plasmodium}, a malarial parasite.} Although Malaria is not life-threatening by its nature, if left untreated, it can cause severe illness and prove to be fatal. The disease was eliminated from American and European continents by first half of twentieth century but is still very common in South Asia and Sub-Saharan Africa. In 2017 alone, there were more than 219 million cases of Malaria and resulted in deaths of more than 435,000 people worldwide \citep{who}. 

In February 2019, a new Malaria vaccine \texttt{RTS,S} - known by the trade name \textit{Mos-Quirix} - was approved for human trials in three countries - Ghana, Malawi and Kenya - coordinated by WHO.  The study is expected to get over by December 2022. However, in last few months, several pharmaceutical majors have begun showing interest in the vaccine's mass production, and the investors want to estimate the coverage ratio - defined by \emph{the vaccine population count divided by the total population}.

{The chapter is outlined as follows. In Section 2, we start with the standard GP model for scalar valued response and present the dynamic SVD-based GP model. Further we discuss the localized dynamic GP model for handling big data. Section 3 explains how dynamic GP model is used for predicting vaccination coverage, with model inputs and built-in R packages. They are illustrated with model outputs on a world map. Finally, concluding remarks and recommendations are suggested in Section 4.}

\section{Evolution of Dynamic GP Model}
In this section, we present a sequence of statistical surrogate models starting from the most basic GP model which emulates deterministic computer simulators  returning scalar outputs, to dynamic GP model that acts as a surrogate to time-series valued simulators. {The models are supported by a brief explanation of their theoretical foundations, an associated example and R implementation.}

\subsection{Basic GP Model}
Gaussian process models are immensely popular in computer experiment literature for emulating computer simulator outputs. In one of the pioneering research, \cite{sacks1989} suggested using realizations of Gaussian stochastic process to model deterministic scalar-valued simulator outputs. However, the notion of such statistical models originate from the kriging literature in Geostatistics.

Let the training data consist of $d$-dimensional input and $1$-dimensional output of the computer simulator, denoted by $x_i=(x_{i1},x_{i2},\dots ,x_{id})$ and $y_i=y(x_i)$, respectively. Then, the GP model is written as
\begin{equation}
y_i = \mu + z(x_i),\quad  i=1,2,\dots,n,
\label{eqn:gpmodel}
\end{equation}
where $\mu$ is the overall mean, and $\{z(x), x\in [0,1]^d\} \sim GP(0, \sigma_z^2R(,))$ with $E(z(x))=0$, $Var(z(x))=\sigma_z^2$, and $Cov(z(x_i),z(x_j)) = \sigma_z^2 R(x_i, x_j)$ where $R(,)$ is a positive definite correlation function. Then,  any finite subset of variables $\{z(x_1), z(x_2), ..., z(x_n) \}$, for $n\ge 1$, will jointly follow multivariate normal distribution. That is, $Y=(y_1, y_2, \dots, y_n)' \sim MVN (\mu {1_n}, \sigma^2_zR_n)$, where $1_n$ is an $n \times 1$ vector of all $1$'s, and $R_n$ is an $n\times n$ correlation matrix with $(i, j)$-th element given by $R(x_i, x_j)$ (see \cite{sacks1989,santner2003,rasmussen2006} for more details).

The model described by (\ref{eqn:gpmodel})  is typically fitted by either maximizing the likelihood or via Bayesian algorithms like Markov chain Monte Carlo (MCMC). As a result, the predicted response $\hat{y}(x_0)$ for an arbitrary input $x_0$ can be obtained as a conditional expectation from the following $(n+1)$-dimensional multivariate normal distribution:
\begin{equation}
\left(
\begin{array}{c}
y(x_0)  \\
Y 
\end{array}
\right)
=
N
\left(
\left(
\begin{array}{c}
\mu  \\
\mu 1_n 
\end{array}
\right)
,
\left(
\begin{array}{cc}
\sigma_z^2 & \sigma_z^2 r'(x_0)  \\
\sigma_z^2 r(x_0) & \sigma_z^2 R_n 
\end{array}
\right)
\right),
\end{equation}
where $r(x_0)=[corr(x_1,x_0), \ldots, corr(x_n,x_0)]'$. The predicted response $\hat{y}(x_0)$  is the same as the conditional mean:
\begin{equation}
E(y(x_0)|Y)=\mu + r(x_0)'R_n^{-1} (Y - 1_n\mu),
\label{eqn:post_mean}
\end{equation}
and the associate prediction uncertainty estimate (denoted by $s^2(x_0)$) can be quantified by the conditional variance:
\begin{equation}
Var(y(x_0)|Y)=\sigma_z^2(1-r'(x_0)R_n^{-1}r(x_0)).
\label{eqn:post_var}
\end{equation}

\begin{figure}[h!]\centering
	\includegraphics[width=4.25in]{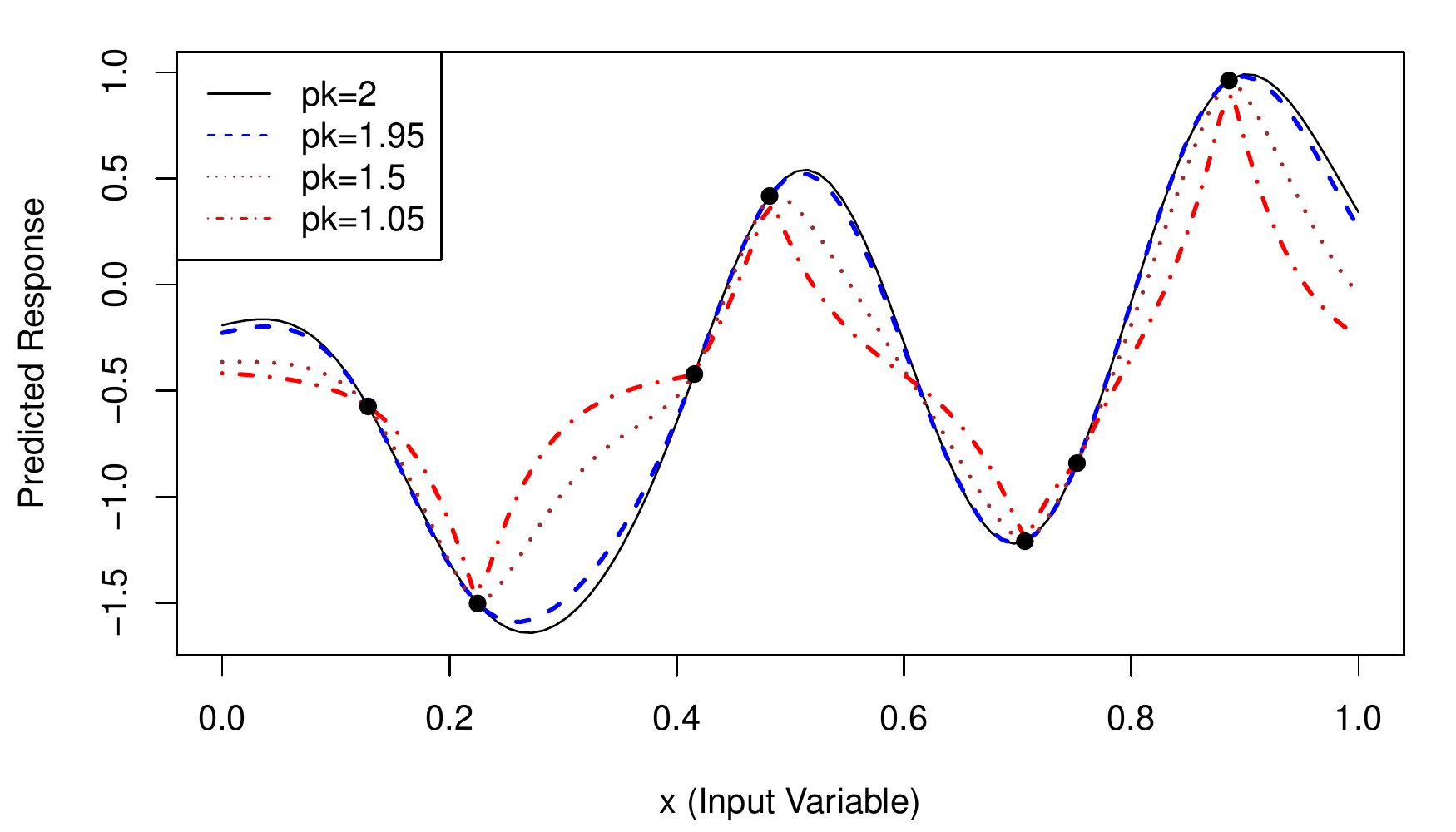}
	\caption{The mean predictions obtained using $\texttt{GPfit}$, when the true simulator response is generated using $f(x)  = ln(x+0.1)+\sin(5 \pi x)$}.
	\label{fig:gpexample1}
\end{figure}

The most crucial component of such a GP model is the spatial correlation structure, $R(,)$, which dictates the `smoothness' of the interpolator that passes through the observations. By definition, any positive definite correlation structure would suffice, but the most popular choice is the power-exponential correlation family given by
\begin{equation}
R(x_i, x_j)  = \prod_{k=1}^d \exp\{-\theta_k |x_{ik}-x_{jk}|^{p_k}\},
\label{eqn:corr}
\end{equation} 
where $\theta_k$ and $p_k$ controls the wobbliness of the surrogate in the $k$-th coordinate.  A special case with $p_k=2$ for all $k=1,2,...,d$, represents the most popular Gaussian correlation also known as radial basis kernel in Machine Learning literature.  Figure~\ref{fig:gpexample1} demonstrates the significance of $p_k$ in the smoothness of the mean prediction.

\textbf{Example~1.} Suppose the simulator output is generated by a one-dimensional test function $f(x)  = ln(x+0.1)+\sin(5 \pi x)$, and $X = \{x_1,..., x_{7}\}$ is a randomly generated training set as per the space-filling Latin hypercube design \citep{mckay1979comparison}. We use an R library called {\bf GPfit} \citep{gpfit} for fitting the model via maximum likelihood approach. Figure~\ref{fig:gpexample1} shows the fitted surrogate along with the true simulator response curves.

Clearly, the choice of $p_k$ in (\ref{eqn:corr}) plays an important role in determining the smoothness of the predictor. It can be noticed from Figure~\ref{fig:gpexample1}, that $p_k=2$ versus $p_k=1.95$ does not make visible difference in terms of smoothness. However, it turns out that by changing the power from $p_k=2$ to $p_k=1.95$, the numerical stability of the correlation matrix inversion can be immensely increased. 

Depending upon the parameter estimation approach used (i.e., maximum likelihood method, empirical Bayesian, or full Bayesian), the prediction uncertainty estimate may vary. For instance, in empirical Bayesian approach, the parameters $\mu, \sigma$ and $\theta$ in $R_n$ are replaced by their maximum a-posteriori (MAP) estimates. On the other hand, the MLE based approach, starts by maximizing the likelihood with respect to $\mu$ and $\sigma^2_z$, giving closed form expressions as
\begin{equation}
\hat{\mu} = (1_n' R_n^{-1} 1_n)^{-1} (1_n' R_n^{-1} Y),
\label{eqn:mu}
\end{equation}
and
\begin{equation}
\hat{\sigma}_z^2 = \frac{(Y-1_n {\mu})' R_n^{-1} (Y-1_n {\mu})}{n},
\label{eqn:sigma2z}
\end{equation}
conditional on the value of $\theta$ in $R_n$.  The hyperparameter $\theta$ is further estimated by maximizing the profile likelihood, which is typically an intensive optimization problem. \cite{sacks1989} reports the prediction uncertainty estimate as
\begin{equation}
s^2(x_0)=\sigma_z^2\left(1-r'(x_0)R_n^{-1}r(x_0) +   \frac{(1-{\bf 1_n}'R_n^{-1}r(x_0))^2}{{\bf 1_n}'R_n^{-1}{\bf 1_n}'}  \right),
\label{eqn:mse_mle}
\end{equation}
which accounts for additional uncertainty due to the prediction of unknown constant mean $\mu$. Of course, the difference between  (\ref{eqn:mse_mle}) and (\ref{eqn:post_var})  can be somewhat substantial. See Example~2 for an illustration using a test function based computer simulator.

\begin{figure}[h!]\centering
	\includegraphics[width=3.05in]{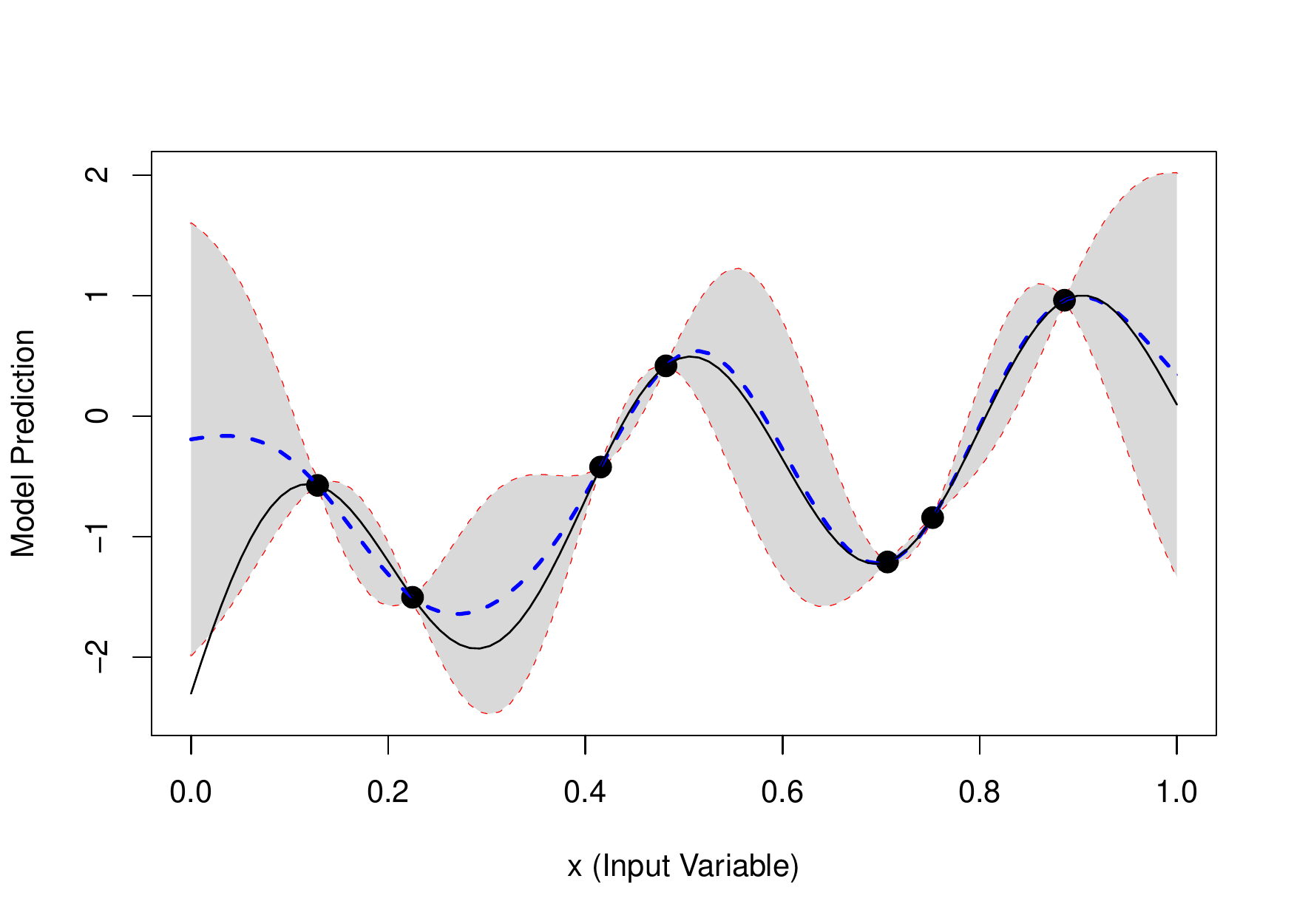}
	\includegraphics[width=3.05in]{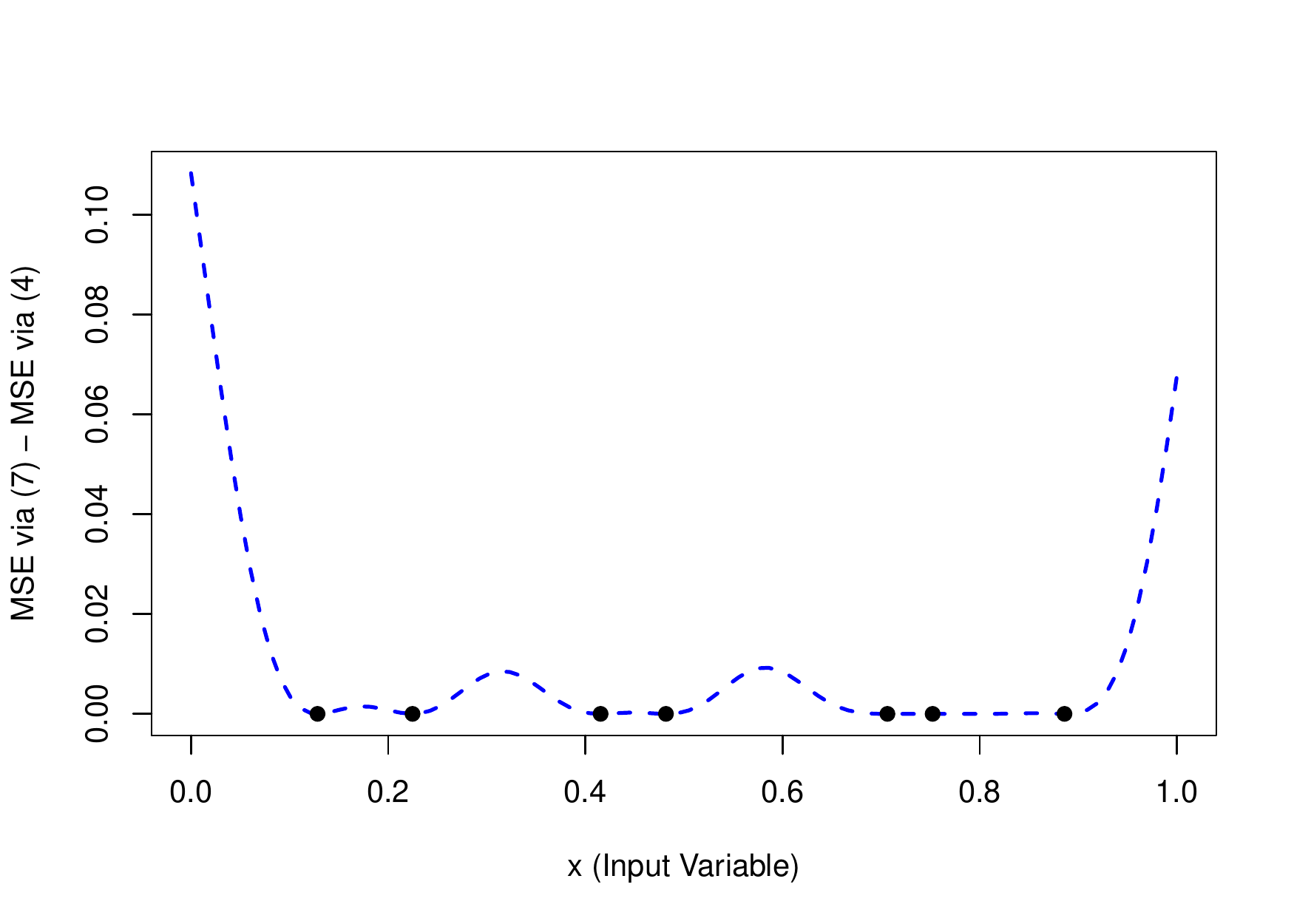}
	\caption{The black solid dots are the training data points. Left panel: The blue dashed curve is the mean prediction obtained using $\texttt{GPfit}$, the  black solid curve is the true simulator response curve $f(x)  = ln(x+0.1)+\sin(5 \pi x)$, and the shaded area represent the uncertainty quantification via $\hat{y}(x) \pm 2s(x)$. Right panel: The prediction uncertainty obtained via MLE as in (\ref{eqn:mse_mle}) -- the posterior variance estimate as in (\ref{eqn:post_var})}.
	\label{fig:gpexample2}
\end{figure}

\textbf{Example~2.} Considering the same setup as in Example~1,  Figure~\ref{fig:gpexample2} shows the fitted surrogate along with the prediction uncertainty estimates obtained via GPfit and the two formulations  (\ref{eqn:post_var}) and (\ref{eqn:mse_mle}).

From the right panel of Figure~\ref{fig:gpexample2}, it is clear that the third term in the prediction uncertainty estimate  (in (\ref{eqn:mse_mle})) is relatively large in the unexplored input regions. As a result, it is recommended to account for uncertainty quantification due to the estimation of unknown model parameters. 

Several additional theoretical and numerical issues on GP models require more careful understanding. See \cite{santner2003, rasmussen2006}, and \cite{harshvardhan2019}, for more details on optimization of likelihood, near-singularity of correlation matrices, choice of correlation kernel, parametrization of hyper-parameters, and the choice of mean function.

%================================
%
%\textcolor{red}{Focus on the power illustration in $R_{ij}$,}
%
%================================

%forretal <- function(x,t,shift=1)
%{
%	par1 <- x[1]*6+4
%	par2 <- x[2]*16+4
%	par3 <- x[3]*6+1
%	t <- t+shift
%	y <- (par1*t-2)^2*sin(par2*t-par3)
%}
%timepoints <- seq(0,1,len=200)
%design <- lhs::randomLHS(100,3)
%test <- lhs::randomLHS(20,3)
%## evaluate the response matrix on the design matrix
%resp <- apply(design,1,forretal,timepoints)

% svdgp, lasvdgp

%======================

\subsection{Dynamic GP Model}

Experimentation via dynamic computer simulators arise in various applications, for
example, rainfall-runoff model (\citet{conti2009gaussian}), vehicle
suspension system (\citet{bayarri2007computer}), and  population growth model for European red mites (\cite{zhang2017local}).  The real-life application presented in this chapter comes from the pharmaceutical industry, where the investors want to predict the coverage of a particular malaria vaccine called RTS,S/AS01 (Mos-Quirix) around the globe over a 20-year window.

The time-series dependence in the simulator response makes the statistical emulation substantially more challenging as compared to the standard GP model presented in the previous section. Recently, a few attempts have been made in this regard. For example, \citet{conti2009gaussian} constructed dynamic emulators by using a one-step transition function of state vectors to emulate the computer model movement from one time step to the next. \citet{liu2009dynamic} proposed time varying autoregression (TVAR) models with GP residuals. \citet{farah2014bayesian} extends the TVAR models in \citet{liu2009dynamic} by including the  input-dependent dynamic regression term. Another clever approach is to represent the time series outputs as linear combinations of a fixed set of basis such as singular vectors (\citet{higdon2008computer}) or wavelet basis (\citet{bayarri2007computer}) and impose GP models on the linear coefficients.  \cite{zhang2017local} further extended the singular value decomposition (SVD) based approach for large-scale data. Next, we discuss the basic version of SVD-based GP model developed by \cite{higdon2008computer}.

Suppose the computer simulator outputs have been collected at 
$N$ design points and stored in the $N\times q$ design matrix $\bm{X}=[\bm{x}_1,\dots,\bm{x}_N]^T$,
and $\bm{Y}=[\bm{y}(\bm{x}_1),\dots,\bm{y}(\bm{x}_N)]$ is the corresponding
$L\times N$ matrix of time series responses. Then the SVD on $\bm{Y}$ gives
$$\bm{Y}=\bm{U}\bm{D}\bm{V}^T,$$
where $\bm{U}=[\bm{u}_1,\dots,\bm{u}_{k}]$ is an $L\times k$
column-orthogonal matrix, $\bm{D}=\text{diag}(d_1,\dots,d_{k})$ is a
$k \times k$ diagonal matrix of singular values sorted in decreasing
order, $\bm{V}$ is an $N\times k$ column-orthogonal matrix of right
singular vectors, and $k=\min\{N,L\}$. \cite{higdon2008computer} suggested modeling the simulator response as 
\begin{align}\label{eq:md}
\bm{y}(\bm{x})=\sum_{i=1}^pc_i(\bm{x})\bm{b}_i+\bm{\epsilon},
\end{align}
where $\bx\in\mathbb{R}^q$, and $\bm{b}_i=d_i\bm{u}_i\in \mathbb{R}^L$, for $i=1,\dots,p$ represent the orthogonal basis. The coefficients $c_i$'s in (\ref{eq:md}) are assumed
to be independent Gaussian processes, i.e.,
$c_i\sim \mathcal{GP}(0,\sigma^2_iK_i(\cdot,\cdot;\btht_i))$ for
$i=1,\dots,p$, where $K_i$'s are correlation functions. We use the
popular anisotropic Gaussian correlation,
$K(\bm{x}_1,\bm{x}_2;\btht_i)=\exp\{-\sum_{j=1}^q\theta_{ij}(x_{1j}-x_{2j})^2\}$. The residual term $\bm{\epsilon}$ in
(\ref{eq:md}) is assumed to be independent $\mathcal{N}(0,\sigma^2\bI_L)$. The number of significant singular values, $p$, in (\ref{eq:md}), is
determined empirically by the cumulative percentage criterion
$p=\min\{m:(\sum_{i=1}^md_i)/(\sum_{i=1}^{k}d_i)>\gamma\}$,
where $\gamma$ is a threshold of the explained variation. 

In this chapter, we discuss the implementation of this so-called svdGP model by   \cite{zhang2017local}. R library called \emph{DynamicGP} \citep{dynamicGP} provides user-friendly functions for quick usage. The most important function is \texttt{svdGP}, and its usage is illustrated as follows:
\begin{verbatim}
      svdGP(design, resp, frac=0.95, nthread=1, clutype="PSOCK", ...)
\end{verbatim}
where \texttt{design} is the input design matrix, \texttt{resp} is the output response matrix, \texttt{frac} specifies $\gamma=95\%$, and \texttt{nthread} and \texttt{clutype} controls the parallelization of the implementation. There are a few additional arguments of \texttt{svdGP()} that accounts for other nuances of the model fitting process.

For all the model parameters in (\ref{eq:md}), \cite{zhang2017local, dynamicGP} used the maximum a posteriori (MAP) values as the plug-in estimates. To obtain the MAP
estimates of process and noise variance parameters, $\sigma^2_i$ and $\sigma^2$, inverse Gamma priors were used, i.e.,
\begin{align*}%\label{eq:prior}
\begin{aligned}
[\sigma_i^2]\sim\text{IG}\left(\frac{\alpha_i}{2},\frac{\beta_i}{2}\right),
i=1,\dots,p,
\end{aligned} \qquad
\begin{aligned} [\sigma^2]\sim \text{IG}\left(\frac{\alpha}{2},\frac{\beta}{2}\right),
\end{aligned}
\end{align*}
and Gamma prior was used for the hyper-parameter $1/\theta_{ij}$ of the correlation function.

\cite{zhang2017local} show that the approximate predictive
distribution for an arbitrary untried
$\bx_0\in\mathbb{R}^q$ is obtained by
\begin{align}\label{eq:final}
\pi(\bm{y}(\bm{x}_0)|\bm{Y})
\approx \pi(\bm{y}(\bm{x}_0)|\bm{V}^*,\hat{\bm{\Theta}},\hat{\sigma}^2) \approx
\mathcal{N}\big(\bm{B}\hat{\bm{c}}(\bm{x}_0|\bV^*,\hat{\bm{\Theta}}),\bB\bm{\Lambda}(\bV^*,\hat{\bm{\Theta}})\bB^T+\hat{\sigma}^2\bI_L\big),
\end{align}
where $\bB= [d_1\bu_1,\dots,d_p\bu_p] = \bm{U}^*\bm{D}^*$, with $\bm{U}^*=[\bm{u}_1,\dots,\bm{u}_p]$, $\bm{D}^*=\text{diag}(d_1,\dots,d_p)$ and $\bV^*=[\bv_1,\dots,\bv_p]^T$, and
$\hbT=\{\hbt_1,\dots,\hbt_p\}$ and $\hat{\sigma}^2$ are the MAP estimates of the
correlation parameters and noise
variance $\sigma^2$, respectively. As shown in \cite{zhang2017local},
\begin{align}\label{eq:resid-var}
  \hbt_i=\underset{\btht_i\in\mathbb{R}^q}{\mathrm{argmax}}\:|\bK_i|^{-1/2}\left(\frac{\beta_i+\psi_i}{2}\right)^{-(\alpha_i+N)/2}\pi(\btht_i),
  \quad \text{and} \quad \hat{\sigma}^2=\frac{1}{NL+\alpha+2}\left(\br^T\br+\beta\right),
\end{align}
where $\bm{K}_i$ is the $N\times N$ correlation matrix on the design
matrix $\bm{X}$ with the $(j,l)$th entry being
$K(\bm{x}_j,\bm{x}_l;\hbt_i)$ for $i=1,\dots,p$ and $j,l = 1,\dots,N$,
$\psi_i=\bv_i^T\bK_i^{-1}\bv_i$, $\pi(\btht_i)$ is the prior
distribution of $\btht_i$ and
$\br=\tvec(\bY)-(I_N\otimes\bB)\tvec(\bV^{*T})$ with $\tvec(\cdot)$
and $\otimes$ being the vectorization operator and the Kronecker
product for matrices, respectively.

The vector of predictive mean of the coefficients at $\bx_0$
is
\begin{eqnarray}\label{eq:coef-pmean}
\hbc(\bx_0,|\bV^*,\hbT) &=&[\hat{c}_1(\bx_0|\bv_1,\hbt_1),\dots,\hat{c}_p(\bx_0|\bv_p,\hbt_p)]^T \\
&=& [\bk_1^T(\bx_0)\bK_1^{-1}\bv_1,\dots,\bk_p^T(\bx_0)\bK_p^{-1}\bv_p]^T, \nonumber
\end{eqnarray}
where
$\bm{k}_i(\bm{x}_0)=[K(\bm{x}_0,\bm{x}_1;\hbt_i),\dots,K(\bm{x}_0,\bm{x}_N;\hbt_i)]^T$.
The predictive variance $\bm{\Lambda}(\bV^*,\hat{\bm{\Theta}})$ of the
coefficients at $\bx_0$ is a $p\times p$ diagonal matrix with the
$i$th diagonal entry being
\begin{align}\label{eq:coef-ps2}
  \hat{\sigma}_i^2(\bm{x}_0|\bv_i,\hat{\bm{\theta}}_i)=\frac{(\beta_i+\bv^T_i\bK_i^{-1}\bv_i)\left(1-\bm{k}^T_i(\bm{x}_0)\bm{K}^{-1}_i\bm{k}_i(\bm{x}_0)\right)}{\alpha_i+N}.
\end{align}

Example~3 illustrates the implementation of svdGP model for a test function based computer simulator model via the R library DynamicGP \citep{dynamicGP}.

\begin{figure}[h!]\centering
	\includegraphics[width=6.0in]{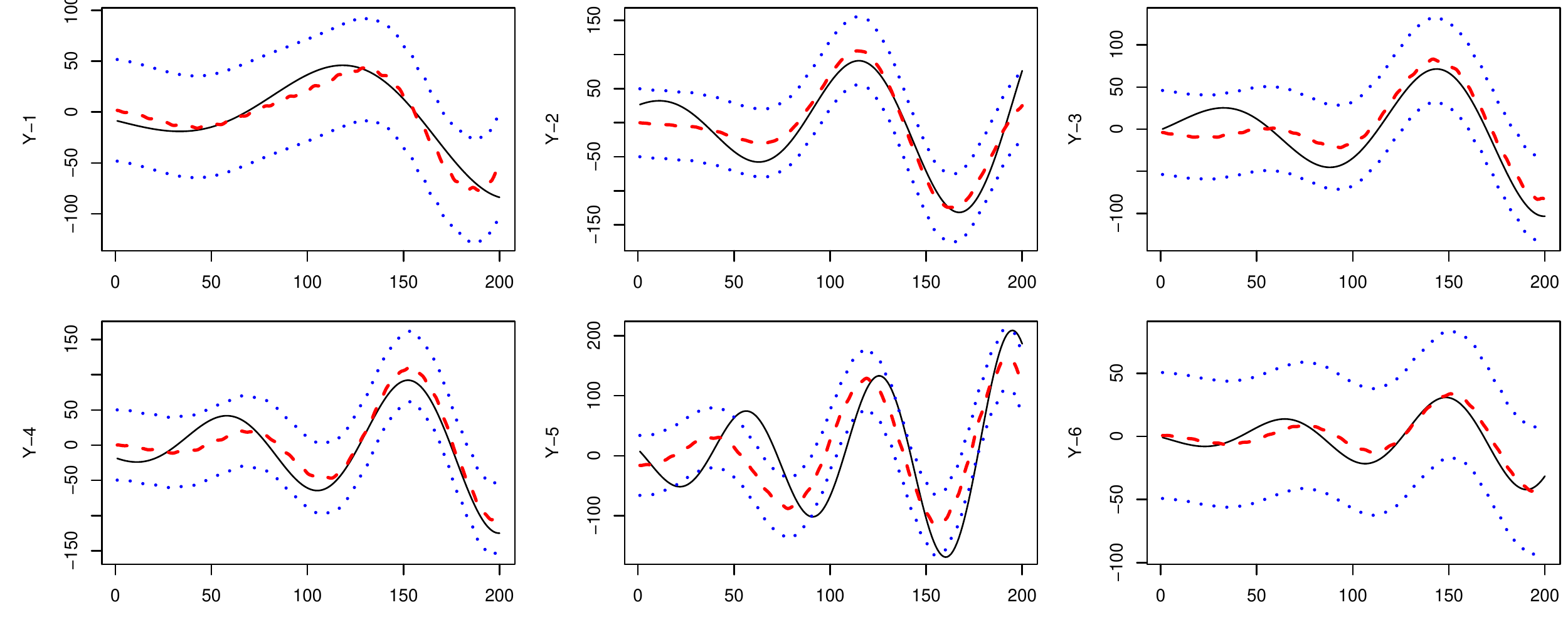}
	\caption{Model prediction for six randomly chosen inputs. Each panel shows the true simulator response (black solid curve), the mean predicted svdGP fit (dashed red curve), and the uncertainty bounds (blue dotted curves).}
	\label{fig:gpexample3}
\end{figure}

\textbf{Example~3.} Suppose the time-series valued response is generated using the following test function \citep{forrester2008} which takes 3-dimensional inputs,
\begin{align}\label{eq:ex2}
f(\bm{x},t) = (x_1t-2)^2\sin(x_2t-x_3),
\end{align}
where $\bm{x}=(x_1,x_2,x_3)^T\in[4,10]\times[4,20]\times[1,7]$, and
$t\in[1,2]$ is on a 200-point equidistant time-grid. We used \texttt{svdGP()} function in the R library  \emph{DynamicGP} for easy implementation. Figure~\ref{fig:gpexample3} illustrates the implementation, by first fitting the svdGP model to a training set of 20 input points randomly generated via maximin Latin hypercube design in the three-dimensional hyper-rectangle $[4,10]\times[4,20]\times[1,7]$, and then predicting the time-series valued simulator output using \texttt{svdGP()} function. 

From Figure~\ref{fig:gpexample3}, it is clear that the fitted surrogate model predictions are reasonable approximations of the simulator outputs at the design points. We fitted svdGP model using the default settings of \emph{DynamicGP} package. Of course, one can play around with other arguments to obtain better (more accurate) predictions.

Both, the basic GP models (in Section~2.1) which emulates scalar-valued simulator outputs, and the svdGP models (in Section~2.2) used for emulating time-series valued dynamic simulator responses, require numerous inversions of $n\times n$ correlation matrices - this is computationally intensive and prohibitive if $N$ (the sample size) is large. For instance, in our motivating application where the training data size is $N=146$ (see Section~3), model fitting via either likelihood method or a Bayesian approach would be computationally burdensome unless the codes are parallelized on heavy computing clusters. The next section briefly reviews GP-based models for large data.

\subsection{Generalizations for Big Data}

Thus far, several techniques have been proposed to account for the large size of the data while building a GP-based surrogate, see \cite{santner2003, harshvardhan2019} for quick reference.  A naive yet popular approach is to fit several local inexpensive (somewhat less accurate) models instead of one big (supposedly more precise) model. The method of searching for local neighborhood can be as simple as finding the \emph{k-nearest neighbours} (k-NN) at the point of prediction. For scalar-valued simulators, \citet{emery2009kriging} built a more efficient local neighbourhood by sequentially including data that make the kriging variance decrease more.  \citet{gramacy2015local}  improved the prediction accuracy by using a greedy algorithm and an optimality criterion for finding a non-trivial local neighborhood set. \cite{zhang2017local} extended this approach further for the svdGP model.

Assuming the total training data size is $N$, and we wish to predict the simulator response at $x_0$. Then, the main idea behind this greedy approach in \cite{gramacy2015local, zhang2017local} is to first use {k-NN} approach for finding $n_0$ neighbours from the training data, and then sequentially obtain the remaining $n-n_0$ points by using an optimality criterion. This proposed greedy-sequential method known as lasvdGP (locally approximate svdGP) is computationally very efficient as compared to the full scalar-GP/svdGP, and much more accurate than the naive k-NN-based svdGP model (referred to as knnsvdGP). 

The following functions in the R library \emph{DynamicGP} can be used for easy implementation:

\begin{verbatim}
   knnsvdGP(design,resp, nn=20, ..., nthread = 1, clutype="PSOCK")
   lasvdGP(design, resp, n0=10, nn=20, ..., nthread = 1, clutype="PSOCK")
\end{verbatim}
where \texttt{design, resp, nthread}, and \texttt{clutype} are the same as in \texttt{svdGP()}, and the important additional parameters are \texttt{nn} - the size of the local neighbourhood set (on which the local GP models have to be built), and \texttt{n0} - size of the local neighbourhood set to be found via k-nearest neighbours which will server as the starting point of the greedy sequential approach for building the local neighbourhood set. 

\begin{figure}[h!]\centering
	\includegraphics[width=6.0in]{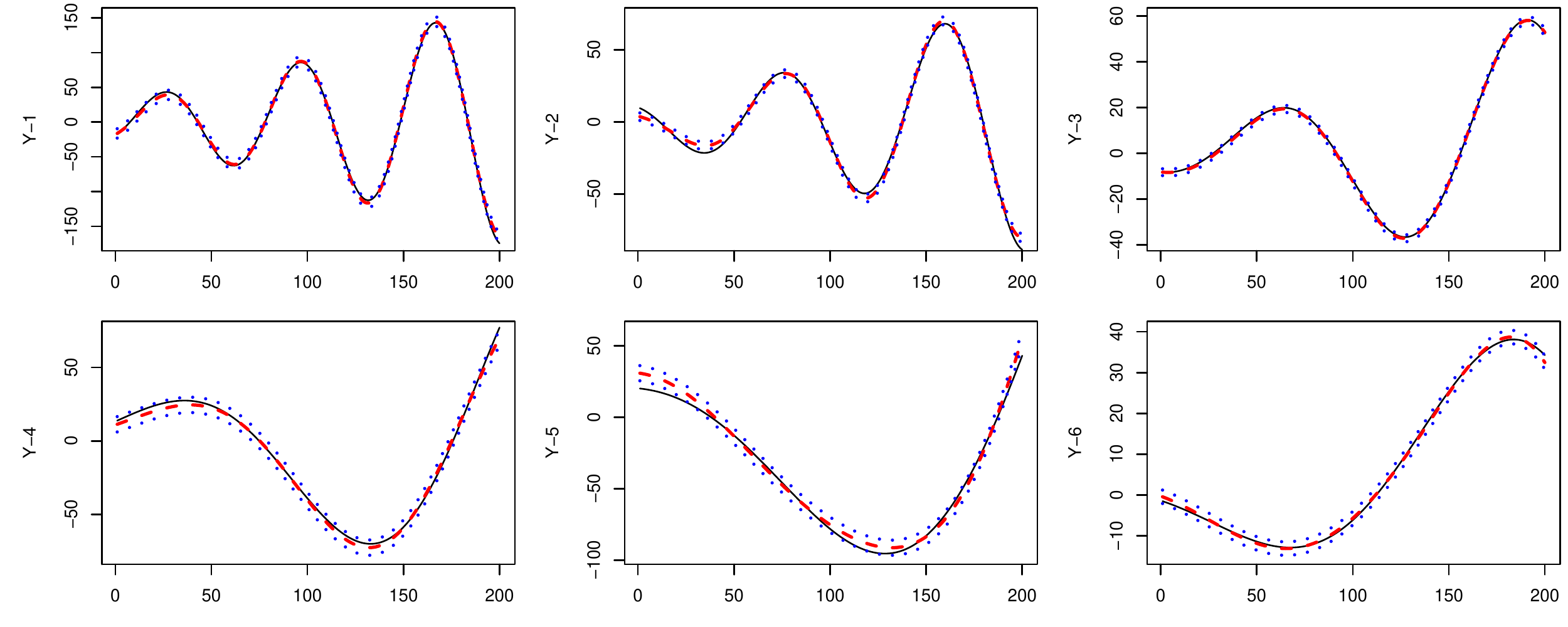}
	\caption{Model prediction for six randomly chosen inputs. Each panel shows the true simulator response (black solid curve), the mean predicted lasvdGP fit (dashed red curve), and the uncertainty bounds (blue dotted curves).}
	\label{fig:lasvdgp-eg4}
\end{figure}

\textbf{Example~4.} Suppose the simulator response is generated using the same test function as in Example~3, but the training data is obtained on a $N=500$-point random Latin-hypercube design in the input space: $[4,10]\times[4,20]\times[1,7]$. In such a case, fitting a full svdGP is certainly infeasible on a regular laptop or desktop. Thus, we  rely on fitting the localized surrogate models like knnsvdGP and lasvdGP. Figure~\ref{fig:lasvdgp-eg4} shows the surrogate fits with $n_0=20$ and $nn=30$ local neighbourhood point sets for  lasvdGP model. 

From Figure~\ref{fig:lasvdgp-eg4}, it is clear that the surrogate fits are much better approximations of the underlying truth (as compared to the illustration in Example~3), which is however expected as the training size is 500 (much bigger than 20 point design in Example~3). Interestingly, the error bounds around the predicted mean response are too narrow and sometimes do not cover the true simulator output. It can perhaps be attributed to the fact that the R library \emph{DynamicGP} uses MAP estimators and not the full Bayesian approach. It is often believed that the latter approach accounts for more uncertainty in the model fitting process.

\section{Application: Malaria Vaccination Coverage}

In this section, we use the historical data on worldwide vaccination coverage for several diseases to predict the coverage ratio of a new Malaria vaccine. The diseased typically experience fevers, chills and flu like illnesses \citep{cdc1} with the symptoms varying in their severity on a case-by-case basis. This can be lethal if not treated properly, and a 2002 study by \cite{greenwood2002} tells us the serious state of Malaria in the world (see \cite{who} for {latest detailed} report).

Recently, a new Malaria vaccine \texttt{RTS,S} (also known as \textit{Mos-Quirix}) has been showing promising results for human trials in Ghana, Malawi and Kenya. Malaria Vaccine Implementation Programme (MVIP), coordinated by WHO, is being funded by a global fund comprising (1) Gavi - The Vaccine Alliance, (2) UNITAID and (3) PATH.  As of now, no results have been made public, and the study is expected to get over by December 2022. However, in last few months, several pharmaceutical majors have begun showing interest in the vaccine's mass production.

Major limitations in the success of a Malaria vaccine are technical and economic feasibility \citep{moorthy2004}. With the current human trials underway, the former is largely solved; however, the latter remains. A study on predicting coverage ratios would immensely benefit to attract global monies -- by corporates and philanthropist funds {-- to the cause}. Recall that the coverage ratio is defined by \emph{the vaccine population count divided by the total population}.  Thus, our objective is to predict the coverage ratio for this Malaria vaccine, using the available data on the coverage ratio of other vaccines. Based on earlier studies on vaccines, the following variables have been identified as predictors:

 \begin{itemize}
 	\item \emph{Dosage number} ($X_1$): The value is $k$, if $k$ doses of the vaccine have already been given.   \cite{luman2005} suggested higher the number of dosages,  lower the chance of completing the entire treatment;
 	
 	\item \emph{Dosage time} ($X_2$): number of months after birth when the first dosage is taken; 0 represents `at birth'. {\cite{luman2005} found vaccines which were given at birth had higher coverage as there is no extra effort needed to come to health centre};
 	
 	\item \emph{Efficacy} ($X_3$): recorded in percentage - ability of the vaccine to actually prevent the disease (see \cite{mclean1995}).  {Vaccination doesn't guarantee prevention, assuming if chances of prevention are better, more people will be vaccinated};
 	
 	\item \emph{Incidence per lac} ($X_4$): it is more likely that the parents will give the vaccine to their children if the occurrence of the disease is high. {When incidences are high, the population is more careful about prevention};
 	
 	\item \emph{Communicable} ($X_5$): binary (0: non-communicable, 1: communicable) - assuming that the fear of contagion may drive the vaccination; 
 	
 	\item \emph{Years active} ($X_6$): how long has the vaccine been around for public use (in years).
 \end{itemize}

\begin{figure}[h!]\centering
	\includegraphics[width=6.5in]{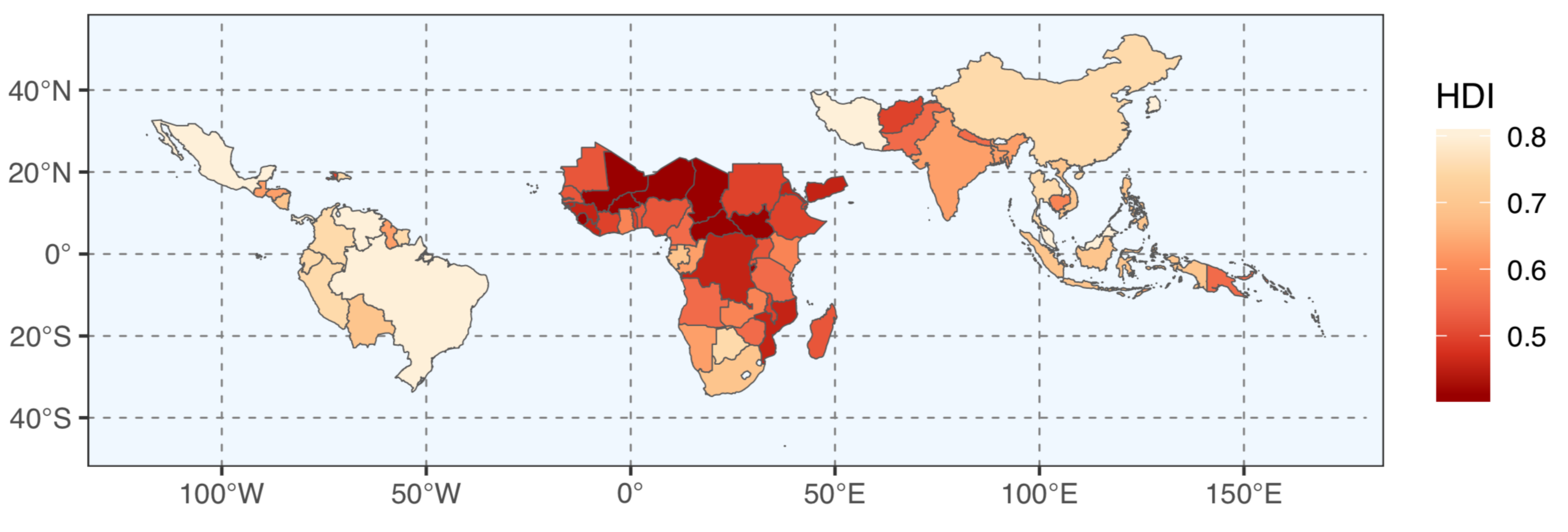}
	\caption{Human Development Index (HDI) value for different countries considered in this study around the globe.}
	\label{fig:hdi_val_map}
\end{figure}

We used data for several vaccines (e.g., Tuberculosis, Diptheria, Hepatitis B, Polio, Japanese Encephalities, Measles, NTetanus, Rubella, and yellow fever) collected on aforementioned variables from {78} countries. We pooled the countries using Human Development Index (HDI) values into {8 groups of size 8, 2 groups of size 7 each}. Figure~\ref{fig:hdi_val_map} depicts the HDI value of different countries.  In total, the data consists of 146 observations - the coverage ratio of different vaccines for 10 country groups observed over 38 year period (from 1980 to 2017),  i.e., $y_{t}(x_i)$, for $t=1,..., 38$ and  the corresponding input $x_i = (x_{i1},x_{i2},...,x_{i7})$, where $i=1,2,...,146$ represent the observation number, $X_1, ..., X_6$ are predictor variables described above, and the seventh input $(X_7)$ is the average HDI value of the country group. 

\begin{figure}[h!]\centering
	\includegraphics[width=6.5in]{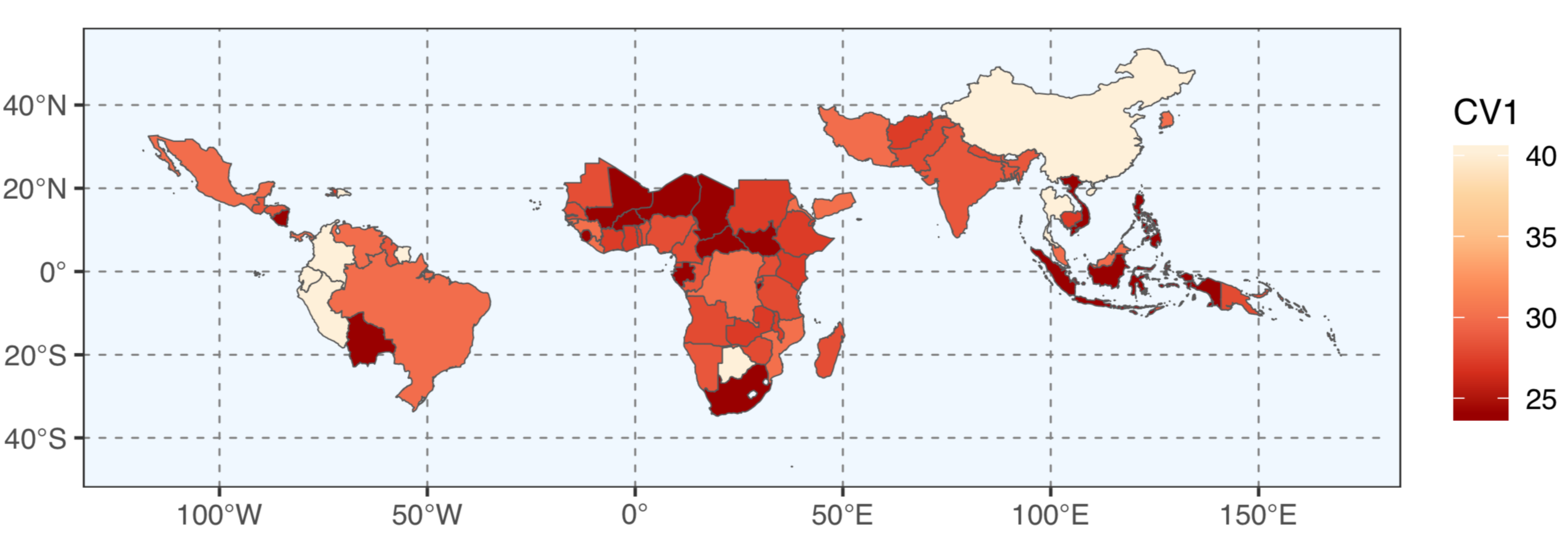}
	\caption{Prediction of coverage ratio for Mos-Quirix at $t=0$ for different country groups using lasvdGP model with $nn=50$ and $n_0=30$ points.}
	\label{fig:world_map_pred_t0}
\end{figure}

\begin{figure}[h!]\centering
	\includegraphics[width=6.5in]{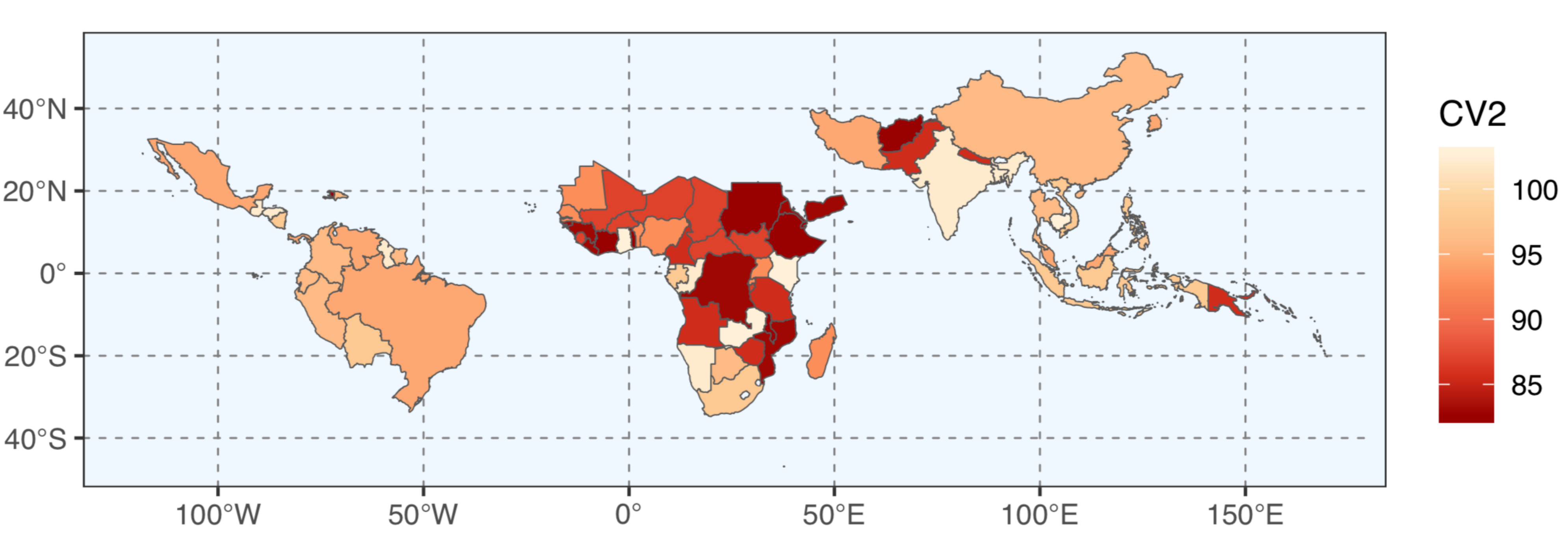}
	\caption{Prediction of coverage ratio for Mos-Quirix at $t=38$ for different country groups using lasvdGP model with $nn=50$ and $n_0=30$ points.}
	\label{fig:world_map_pred_tn}
\end{figure}

Since the training data size is too big to fit a full svdGP model on a standard laptop, we implement the localized model (i.e., lasvdGP model) developed by \cite{zhang2017local} for the model fitting. For a quick illustration, we predict the coverage ratio of the proposed Malaria vaccine {\textit{Mos-Quirix}} for the first dose ($X_1=0$) given to a 6-month old child ($X_2=6$), assuming the disease is not communicable ($X_5=0$) and the vaccine has been around since 1980 (the study period). We run the model with the average observed value of the incidence ($X_4=60$) and a conservative efficacy ($X_3=70$) as compared to other vaccines. We vary the value of $X_7$ for predicting the coverage ratio of Mos-Quirix at $t=0$ and $t=38$ for different country group, see Figure~\ref{fig:world_map_pred_t0} and Figure~\ref{fig:world_map_pred_tn} respectively. 

Note that the development of an accurate model for predicting the coverage ratio is beyond the scope of this chapter. Our main objective is to illustrate the usage of lasvdGP model in a complex real-life statistical problem. Although the overall pattern between Figure~\ref{fig:hdi_val_map} and Figures~\ref{fig:world_map_pred_t0}, \ref{fig:world_map_pred_tn} show positive association among HDI value and coverage ratio, more conclusive remarks require extensive modeling and analysis. One should also look at the dependence with respect to other predictor variables.

\begin{figure}[h!]\centering
	\includegraphics[width=5.5in]{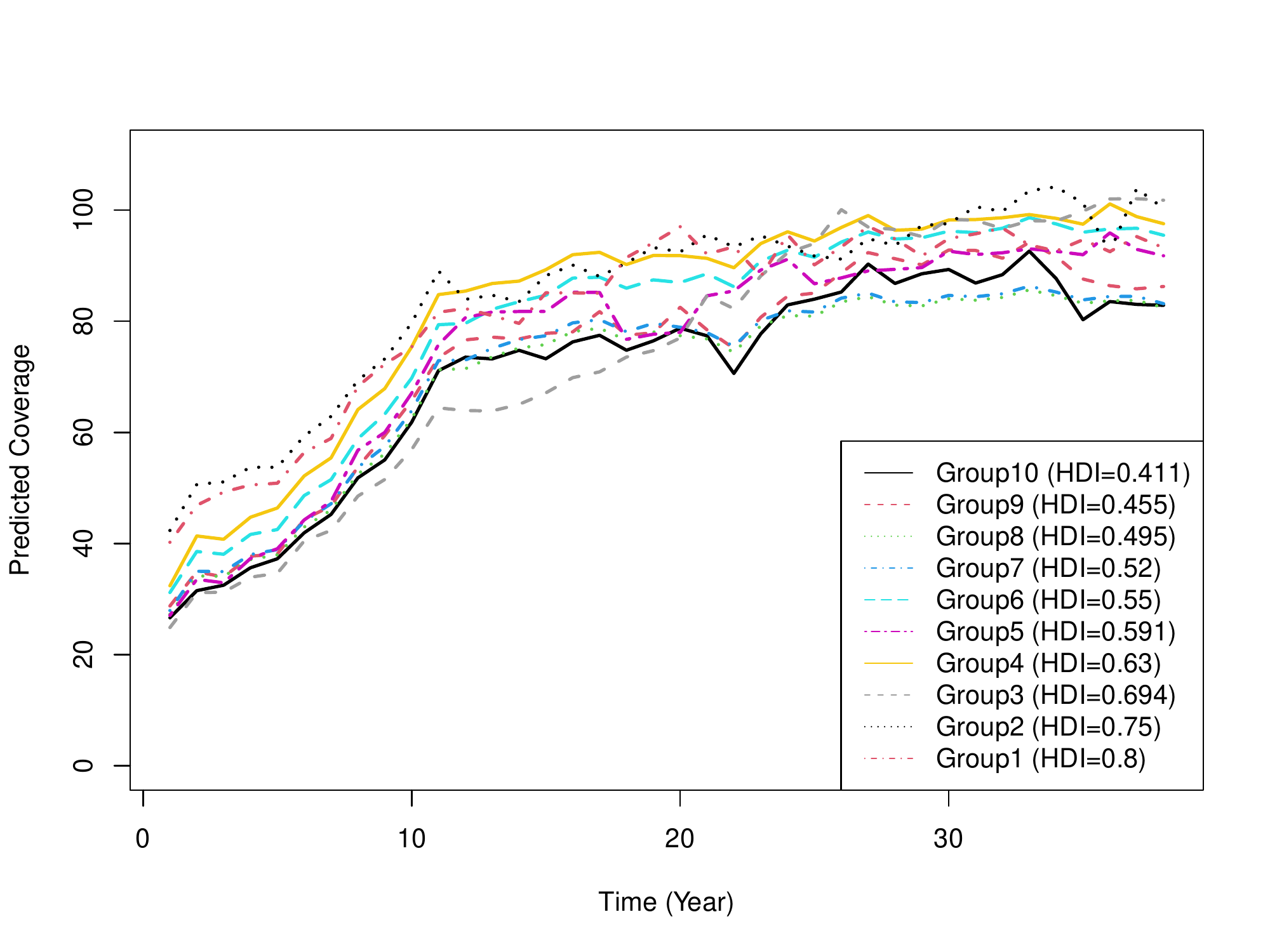}
	\caption{Mean predictions of the Mos-Quirix coverage ratio over time for different country groups classified based on HDI values.}
	\label{fig:hdi_075_pred}
\end{figure}

Figure~\ref{fig:hdi_075_pred} shows the predicted coverage ratio over time - the direct output of lasvdGP model for different country groups.

Clearly, the coverage ratio increases to 100\%. This is expected from this model, but an in-depth analysis is required for more meaningful inference.

%

%======================

\section{Concluding Remarks}
In this chapter, we talked about the popular Gaussian process models and its importance in computer aided experiments for emulating real world phenomena. We discussed various fundamental concepts that drive Gaussian process models, and the statistical interpretations and usages. These models, however, suffer from computational instability due to a variety of reasons, major ones being related to the near-singularity and the cost of inverting correlation matrices. Due to the computational overload, the process is expensive for numerous evaluations, which are needed for parameter estimation. Under the umbrella of big data, we present efficient localized GP models for emulating dynamic (time-series valued) computer simulators.

{The concepts and R implementations are illustrated via several test functions. Finally, we presented an elaborate case-study of how a new Malaria vaccine coverage can be predicted using the dynamic SVD-based GP model. Of course, this is just an illustration and not an attempt to accurately solve the case-study. 
An elaborated second-level modeling and analysis is required to understand how and why the coverage ratios of Mos-Quirix would vary for different countries.}

{One could consider alternative approaches in predicting the coverage ratios. For example, clustering techniques to distribute the countries through their holistic characteristics instead of artificially binning into groups using HDI. One could also simply use a time-series modelling through AR, MA, ARIMA, etc. to predict coverage ratios. }

\section*{Acknowledgements}
{We would like to thank Pradeep Charan, an IPM student at IIM Indore, for brainstorming on the applications of dynamic GP and help in collecting country-level data. We would also like to thank Aspect Ratio (http://aspectratio.in) for stimulating and inspiring this research. The authors also thank the editor and two reviewers for their helpful comments which led to significant revision of the chapter.}
%*************

\bibliographystyle{apalike}
\bibliography{GPbib.bib}

\newpage
\section*{Appendix: R Codes}

The following R code generates the prediction curves in Figure~1 of Example~1.  One can change the "power" argument in \verb"GP_fit" and "predict" to fit GP model with different power exponential correlation structures.

\begin{verbatim}
#-------------------------------------------
n = 7; d = 1;
computer_simulator <- function(x) {
   y <-  log(x+0.1)+sin(5*pi*x)
   return(y)
}

set.seed(1)
library(lhs)
library(GPfit)

x = maximinLHS(n,d)
y = computer_simulator(x)

xpred = seq(0,1,length=100)
ytrue = computer_simulator(xpred)

GPmodel = GP_fit(x,y, corr = list(type="exponential", power=1.95))
pred=predict(GPmodel,xnew=xpred, corr = list(type="exponential", power=1.95))
yhat = pred$Y_hat
#-------------------------------------------
\end{verbatim}

The following R code generates the prediction curves in Figure~3 of Example~3.  \verb"ret$pmean[,i]" contains the predicted mean response for the $i$-th input and \verb" ret$ps2[,i]" contains the corresponding mean square error estimates. 

\begin{verbatim}
#-------------------------------------------
set.seed(1234568)

library("lhs")
library(DynamicGP)
 
forretal <- function(x,t,shift=1)
{
    par1 <- x[1]*6+4
    par2 <- x[2]*16+4
    par3 <- x[3]*6+1
    t <- t+shift
    y <- (par1*t-2)^2*sin(par2*t-par3)
}
timepoints <- seq(0,1,len=200)

train <- maximinLHS(20,3)
resp <- apply(train,1,forretal,timepoints)
test <- randomLHS(50,3)

ret <- svdGP(train,resp,test,nstarts=5)
#-------------------------------------------
\end{verbatim}

For generating the predictions in Figure~4 of Example~4, we only need to replace the last line of the previous code ('\verb"ret <- svdGP(...)"') with the following code.

\begin{verbatim}
#-------------------------------------------
retl <- lasvdGP(atrain,resp,atrain,nn=30,n0=20,nstarts=5)
#-------------------------------------------
\end{verbatim}

\end{document}